\documentclass[twocolumn,showpacs,prl,amsmath]{revtex4}
\usepackage{bm}

\usepackage{epsfig}
\newcommand{\nix}[1]{}
\usepackage{color}
\usepackage{graphics}
\usepackage{amsmath}
\usepackage{amssymb}

\begin{document}

\title{Cyclotron-resonance-assisted photon drag effect in InSb/InAlSb quantum wells excited by terahertz radiation
}

\author{ S~Stachel,$^1$ G\,V~Budkin,$^{2}$ U~Hagner,$^1$ V\,V~Bel'kov,$^{2}$  
M\,M~Glazov,$^{2}$ S\,A~Tarasenko,$^{2}$  S\,K~Clowes,$^{3}$ T~Ashley,$^{4}$ 
A\,M~Gilbertson,$^{5}$ and S\,D~Ganichev$^{1}$}

\affiliation{$^1$Terahertz Center, University of Regensburg, 93040
Regensburg, Germany} 
\affiliation{$^2$Ioffe Physical-Technical
Institute, Russian Academy of Sciences, 194021 St.\,Petersburg,
Russia} 
\affiliation{$^3$Advanced Technology Institute and SEPNet, University
of Surrey, Surrey GU2 7XH, UK} 
\affiliation{$^4$ School of Engineering, University of Warwick, Coventry CV4 7AL, UK}
\affiliation{$^5$ Blackett Laboratory, Imperial College, London SW7 2BZ, UK}

\begin{abstract}
We report on the observation of the cyclotron-resonance-assisted photon 
drag effect. Resonant photocurrent is detected in InSb/InAlSb quantum wells 
structures subjected to a static magnetic field and excited
by terahertz radiation at oblique incidence. The developed theory based 
on Boltzmann's kinetic equation is in a good agreement with the experimental findings. 
We show that the resonant photocurrent originates 
from the transfer of photon momentum to free electrons drastically enhanced
at cyclotron resonance. 
\end{abstract}
\maketitle

\section{Introduction}

Indium antimonide based quantum well (QW) structures exhibiting 
the highest room temperature electron mobility 
of any semiconductors are considered as a prospective material 
for the next generation of electronic devices
being of particular interest for such applications 
as high mobility transistors~\cite{ashley2007} and ballistic 
transport based electronics.~\cite{gilbertson2011PRB,gilbertson2011Appl,Goel2004} 
Apart from the high mobility  caused by the small effective electron mass, 
these nanostructures are also characterized by  large dielectric constant,
enhanced Land$\acute{e}$ $g$-factor and a 
strong spin-orbit coupling,~\cite{gilbertson2008,gilbertson2009,khodaparast2004} 
making them particularly interesting for  the studies of spin physics.
Whereas electronic transport in InSb/AlInSb heterostructures has been 
investigated in depth, less is known about their optoelectronic properties.
At the same time, already the first experiments  have  
demonstrated that the photocurrents excited by 
infrared/terahertz radiation give an access
to the Rashba spin-splitting of 
the conduction band~\cite{Khodaparast}, 
results in a spin-dependent electron transport~\cite{Stachel2012,Li1,Li2},
prove the high polarization-dependent spin 
susceptibility and enhanced electron-electron 
exchange interaction~\cite{Stachel2012}.
Photoelectric effects have been also shown 
to be an effective way to  inject  spin polarized electric currents and to study 
the anisotropy of the band spin-splitting~\cite{IvchenkoGanichev}.
Being caused by the absorption of electromagnetic radiation, 
photocurrent generation efficiency can be strongly enhanced under  resonant absorption 
conditions. 
In particular, cyclotron resonance (CR) serves as an effective tool to magnify the photoresponse. 
Textbook CR experiments on bulk InSb crystals 
excited by terahertz radiation\cite{Seeger} 
and more recent results on InSb-based QWs\cite{Orr,Khodaparast2003,Gouider}
reveal that the radiation absorption can be increased tenfold
and CR, due to very small effective mass, emerges at low magnetic field of a few Tesla.

Here, we report on the observation and detailed study of the 
cyclotron-resonance-assisted photocurrent induced by terahertz 
laser radiation in InSb/AlInSb QWs. 
We show that, in (001)-grown QWs, the photocurrent 
can be excited at oblique incidence only. Applying radiation
with the frequency  of about 2.5\,THz, the photocurrent 
enhancement by more than 50~times is detected at the 
magnetic field of about 2\,T as compared to the photocurrent excited at zero field. 
Both the resonance position and the enhancement factor correlate well with those of
the optical transmission measurements clearly indicating the CR absorption. 
The developed theory, being in a good agreement 
with experimental findings, demonstrates that the electric current originates 
from the transfer of photon momentum to free electrons, that is the photon-drag effect.

\section{Samples and experimental techniques}
\label{technique}

The experiments are carried out on 
30\,nm wide InSb QW structures grown on (001)-oriented 
semi-insulting GaAs substrates by molecular beam epitaxy. The lower barrier of the 
well consists of In$_{0.9}$Al$_{0.1}$Sb while the top barrier is In$_{0.85}$Al$_{0.15}$Sb 
containing a Te $\delta$-doping layer. The QW hosts a two-dimensional electron gas 
(2DEG) with a carrier density of $n_{s}$ = 5 $\times$ 10$^{11}$\,cm$^{-2}$ and a 
mobility of $\mu$ = 1 $\times$ 10$^{5}$ cm$^{2}$/Vs at 77\,K. The samples are
square shaped with a length of 5~mm. The
sample edges are oriented along 
the $x\parallel [1\bar{1}0]$  and  $y\parallel [110]$  
crystallographic axes. Four ohmic contacts fabricated at the middle of 
each sample edge allow us to probe the photocurrent in 
$x$- and $y$-directions (see the inset in Fig.~\ref{cwtransverse}).

To generate photocurrents in unbiased samples we applied
$cw$ or pulsed terahertz (THz) laser systems. 
As a source of $cw$ radiation, a CH$_{3}$OH laser~\cite{PhysicaE2008} operating at the frequency 
2.54\,THz (photon energies $\hbar \omega = 10.4$\,meV)   was  used. 
The incident power $P\,\approx\,0.5$\,mW at the sample was
modulated at 800\,Hz  by an optical chopper. 
The radiation at oblique incidence was focused in a spot of 
about 1.5\,mm diameter at the center of sample.  The spatial beam 
distribution had an almost Gaussian profile which was 
measured by a pyroelectric camera.\cite{ch1Ziemann2000p3843}
The electric field  $\bm{E}$ of linearly polarized laser radiation was 
oriented either parallel ($p$-polarization) or perpendicularly 
($s$-polarization) to the plane of incidence. Note that in the latter case, 
the electric field vector lied in the QW plane. 
For the pulsed radiation, an NH$_{3}$ laser\cite{JETP1982,DX} 
operating at the frequencies $f$ = 1.07, 2.03, and 3.31\,THz 
($\hbar \omega = 4.4$, 
8.4, and 13.7\,meV, respectively) was applied. 
The laser generates single pulses with a duration of about 100\,ns, peak power 
of $P \approx 5$\,kW, and a repetition rate of 1\,Hz. The radiation power was 
controlled by the THz photon drag detector.\cite{Ganichev84p20}
A typical spot diameter was from 1 to 3\,mm. The beam 
had an almost Gaussian shape, which was 
measured by the pyroelectric camera.

The geometry of the experiments is sketched in the insets in 
Figs.~\ref{cwtransverse} and~\ref{cwparallel}.  
  Samples were excited by 
the laser beam lying in the ($xz$) plane at the angle of incidence $\theta_0$, varied between 
$-30^\circ$ to +30$^\circ$, with respect to the layer normal $z \parallel [001]$.
Both the current components perpendicular, $j_y$,
and parallel, $j_x$, to the plane of incidence 
were investigated. 
The corresponding photocurrents  were measured 
by the voltage drops, $U_{x(y)}$, picked up across a 1\,M$\Omega$ ($cw$ measurements) or 50~$\Omega$ 
(pulsed measurements) load resistors. To record the signal 
in both cases, a lock-in 
amplifier and digital oscilloscope were applied, respectively.
Complimentary transmission measurements were performed applying 
$cw$ laser operating at the frequencies $f$ = 1.63 or 2.54\,THz 
and the Golay cell as a sensitive radiation detector. 
These measurements were done at normal incidence of radiation using linearly polarized as well as 
right-handed ($\sigma^+$) and left-handed ($\sigma^-$) circularly polarized light 
obtained by $\lambda/4$-plates.
Samples were mounted in a temperature variable magneto-optical cryostat. 
Experiments  were  carried out in the 
temperature range between 4.2 and 120\,K and the magnetic field up to 7\,T  applied 
along $z$-direction.

\section{Results}
\label{results}

\begin{figure}[h]    
\includegraphics[width=1.0\linewidth]{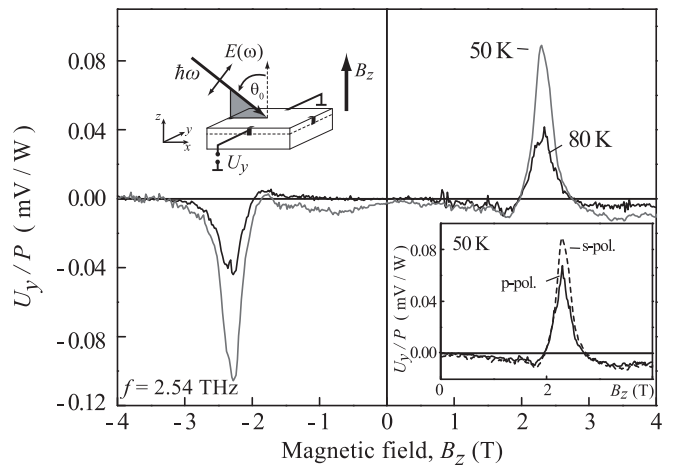}
\caption{
Photosignal $U_{y}$ normalized by the radiation power $P$ 
excited by linearly polarized light ($s$-polarization) with $f$ = 2.54~THz 
as a function of the magnetic field $B_z$, incidence angle $\theta_0 = 20^0$.
The left inset shows the experimental geometry. 
The right one presents the field dependences for $s$- and 
$p$-polarized radiation.
}
\label{cwtransverse}
\end{figure}

\begin{figure}[h]   
\includegraphics[width=1.0\linewidth]{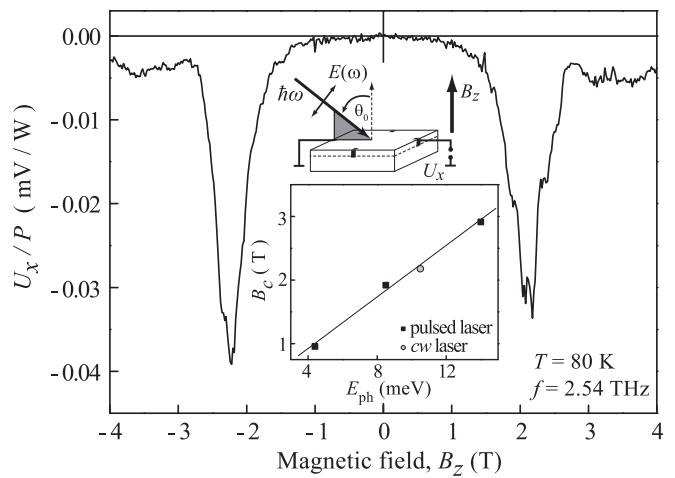}
\caption{
Photosignal $U_{x}$ normalized by the radiation power $P$ 
excited by linearly polarized ($p$-polarization) light with $f = 2.54$~THz 
as a function of the magnetic field $B_z$, incidence angle $\theta_0 = 20^0$.
The top inset shows the experimental geometry. The bottom one demonstrates
 magnetic field positions, $B_{c}$, of the resonant photosignals 
induced by the $cw$ ($f = 2.54$\,THz) and pulsed laser systems as 
a function of the photon  energy $E_{ph}$ obtained at $T = 4.2$\,K.  
}
\label{cwparallel}
\end{figure}

We start with the data obtained at rather high temperatures (above $\approx$~20\,K) 
and applying low power radiation of \textit{cw} laser operating at $f = 2.54$\,THz.
Exciting the sample with  linearly polarized radiation 
 at oblique incidence\cite{footnote1}
and  sweeping  magnetic field, we observed  a resonant photosignal 
with the maximum at $|B_c| = 2.3$\,T, see Figs.~\ref{cwtransverse} 
and~\ref{cwparallel}.
The resonant photoresponses are detected in both \textit{perpendicular} 
and \textit{parallel}
to the light incident plane directions. However, the signal parities are different:
In the former case ($U_y$) it is \textit{odd} (Fig.~\ref{cwtransverse}) 
while in the latter  case  ($U_x$)  it is 
\textit{even} (Fig.~\ref{cwparallel}) 
in the magnetic field $B_z$. 
As an important result, the magnitude of the resonances 
is almost independent of 
the orientation of the radiation electric field vector 
$\bm{E}_{\omega}$  with  respect to 
QW plane. The photocurrents excited by the radiation polarized
in the QW plane ($s$-polarization, $\bm E_{\omega} \parallel y$) 
and the field with the orthogonal polarization  ($p$-polarization,  $E_{\omega,z} \neq 0$) 
are shown in the inset in Fig.~\ref{cwtransverse}.

\begin{figure}[h]  
\includegraphics[width=1.0\linewidth]{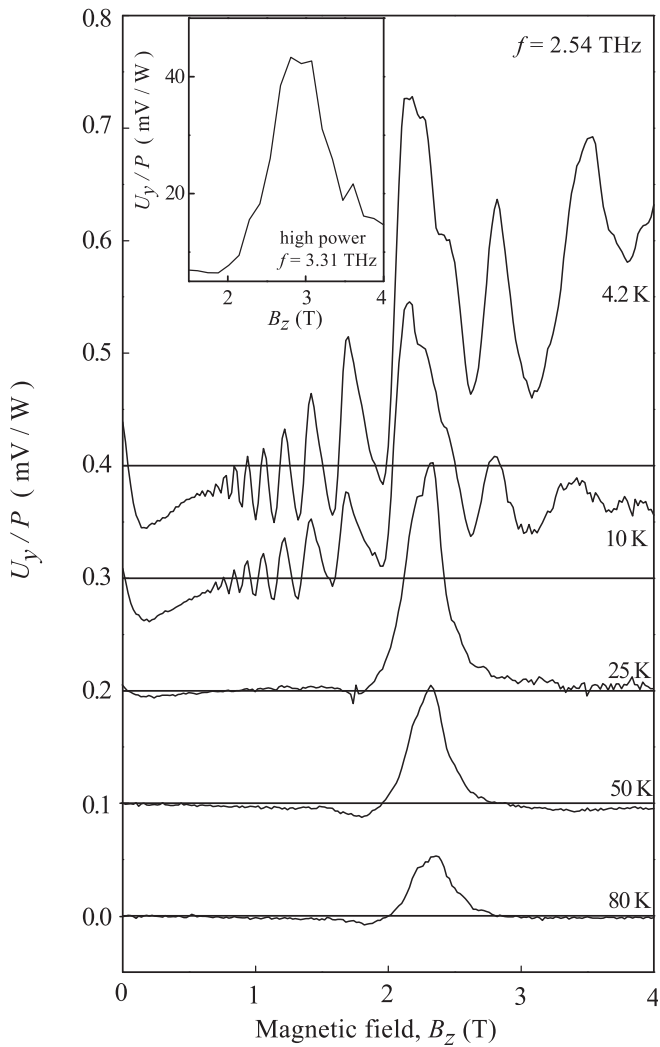}
\caption{Magnetic field dependence of $U_{y}/P$ for linearly 
polarized radiation of  $f = 2.54$\,THz and different 
temperatures $T$. 
The data for $T < 80$~K are consistently shifted by +0.1\,mV/W for each 
step in temperature. 
Inset shows photovoltage resonance observed at 4.2\,K under high power 
excitation of the pulse laser, $f = 3.31$\,THz. The signal $U_y$ in the inset is 
normalized by the maximum detected signals, $U_{y,max}$. 
}
\label{cwdifftemp}
\end{figure}

\begin{figure}[h]   
\includegraphics[width=1.0\linewidth]{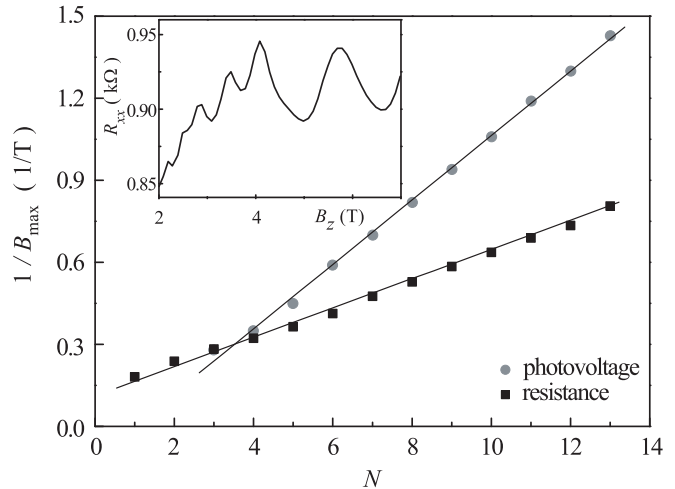}
\caption{
Oscillation maxima of the photocurrent and SdH oscillations measured at $T=4.2$\,K
are assigned an integer index $N$. 
Their 1/$B$-field positions are plotted versus the index.
The inset shows the longitudinal resistance $R_{xx}(B_z)$ measured in the van der Pauw 
geometry for a constant current of 1\,$\mu$A. Note, that the SdH oscillations are 
superimposed by a high background signal resulting from the 
parallel conductance of the $\delta$-doping layer in the AlInSb 
barrier which is well known for InSb/AlInSb 
QWs.\protect\cite{Pooley2010,Gilbertson2011}
}
\label{SdH}
\end{figure}

In the temperature range between 20 and 100\,K, the 
resonances are well described by the Lorentz functions, whose magnitude 
increases with a decrease of the temperature.
For 
lower temperatures, $T < $~20\,K, the magnetic field 
behavior of the photocurrent 
becomes more complicated. Figure~\ref{cwdifftemp} shows
that now the photosignal oscillates upon 
variation of magnetic field. The amplitude of the oscillations 
increases with rising $B_z$ achieves a maximum at $B_z\approx B_c$ and decreases 
for higher magnetic fields. Analysis of the data reveals that the oscillations are
periodic with $1/B_z$.
This is shown in Fig.~\ref{SdH} demonstrating a linear
dependence of the inverse magnetic field positions, $1/B_{max}$, 
of the photosignal maxima  on an integer index $N$. 
Additional magneto-transport measurements demonstrate that the 
period of the observed oscillations is  twice as
that of the Shubnikov-de Haas oscillations (SdH), see Fig.~\ref{SdH}.
At substantially higher radiation intensities 
obtained applying pulsed terahertz laser, the oscillations vanish
even at low sample temperatures, see inset in Fig.~\ref{cwdifftemp}.

\begin{figure}[h]   
\includegraphics[width=1.0\linewidth]{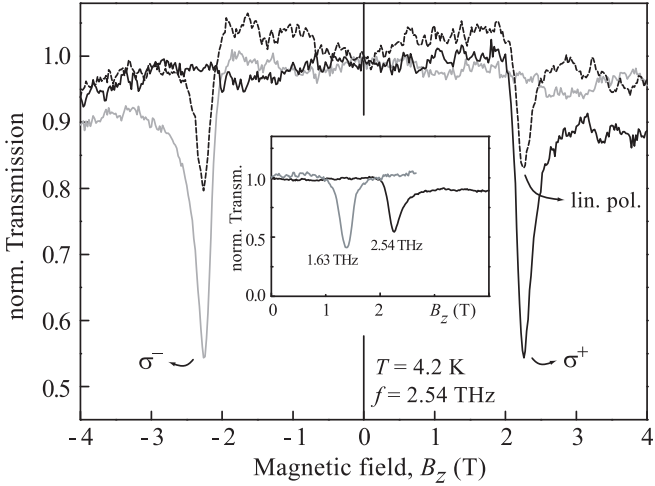}
\caption{Magnetic field dependence of the normalized transmission 
for right and left circularly as well as linearly polarized radiation 
at normal incidence of light. The transmitted radiation power is normalized 
by that at zero magnetic field. 
The data are obtained for $T = 4.2$\,K. The inset shows transmission curves for 
two radiation frequencies of the $cw$  laser. 
}
\label{Transmission}
\end{figure}

The magnetic field $B_c$ corresponding to the resonance 
versus the photon energy are shown in 
the inset in Fig.~\ref{cwparallel}.
The linear scaling of $B_c$ with the radiation frequency, almost Lorentz 
shape of the resonance, and the magnitudes of $B_c$
indicate that the resonances are caused by the cyclotron 
absorption of radiation. 
This conclusion is supported by the magnetic field dependence of the radiation 
transmission measured for normally incident radiation of the $cw$ laser. 
Figure~\ref{Transmission}
shows  that the data obtained for linearly and circularly polarized 
radiation with $f = 2.54$\,THz have a characteristic cyclotron 
resonance  behavior: 
While for left-handed (right-handed) polarized light the transmission shows a sharp 
decrease  at $B_{c} \approx$ -2.3\,T (+2.3\,T) only for one magnetic 
field polarity,
for linear polarization the dips in the transmission are observed 
for both magnetic field 
 polarities and the  dip 
magnitudes are  reduced by the factor of two. 
Additional measurements applying  $cw$ laser radiation with $f=1.63$\,THz prove
that the resonance position shifts linearly with the radiation frequency as it is 
expected for the cyclotron resonance, see the inset in Fig.~\ref{Transmission}. 
The effective mass calculated from the CR is $m^* = 0.025 \cdot m_0$ 
being in a good agreement with the literature data on similar 
structures.\cite{Orr,Nedniyom,Khodaparast2003} 
Momentum relaxation time determined from the resonance width ($1.1 \cdot 10^{-12}$\,s)
is found to be close to that obtained from the transport measurement 
also carried out at 4.2\,K and yielding $\tau_p = 1.3 \cdot 10^{-12}$\,s.

\section{Symmetry analysis}

Photocurrents induced by homogeneous excitation of  QW structures 
with terahertz radiation  
can generally be caused by photogalvanic effects (PGE), resulting
from an asymmetry of the carrier photoexcitation/scattering 
in $\bm k$-space, 
or photon drag  effects (PDE), originating from the transfer of the photon 
momentum $\bm q$ to free carriers (see e.g. \cite{Ivchenko05,GanichevPrettl,PGE3authors}).
We start with the symmetry analysis 
of both effects and demonstrate that the observation of the photocurrent 
for $s$-polarized radiation allows us to exclude PGE as a possible origin 
of the photocurrent and, consequently, indicates that the current is caused 
by the photon drag effect. 
PGE and PDE are phenomenologically described by  equation which couples 
\textit{dc} electric current with the amplitude and wave vector of the electromagnetic 
field inside the medium.
To second order in the field amplitude (linear regime in the 
radiation intensity) and first order in the wave vector,
such equation has the form\cite{Ivchenko05,GanichevPrettl}
\begin{equation}
\label{current_general}
j_{\alpha}=\sum\limits_{\beta \gamma}\chi_{\alpha \beta \gamma}(B) E_{\omega,\beta} E_{\omega,\gamma}^*+
\sum\limits_{\beta \gamma\delta}\phi_{\alpha \beta \gamma \delta} (B)\: q_{\beta} E_{\omega,\gamma} E_{\omega,\delta}^* \:,
\end{equation}
where $\bm{j}$ is current density, $\bm{E}_{\omega}$ is the complex amplitude of the electric field of THz wave
\begin{equation}
\bm E(\bm{r},t)=\bm E_{\omega} {\rm e}^{i\bm q \cdot \bm r -i \omega t} + \bm E_{\omega}^*{\rm e}^{-i\bm q \cdot \bm r +i \omega t}\:,
\end{equation}
$\bm{q}$ is the photon wave vector, and the indices 
$\alpha$, $\beta$, $\gamma$, $\delta$ enumerate the Cartesian components.
The third rank tensor $\chi_{\alpha\beta\gamma}$ and the forth rank tensor $\phi_{\alpha\beta\gamma\delta}$ 
describe the  photogalvanic effects 
and  photon drag effects, respectively. Their components may depend 
on static magnetic field $\bm B$. 

Our experiments are carried out on (001)-grown quantum wells 
which are described by the C$_{2v}$ point group. 
For these structures subjected to the magnetic field $\bm{B}$ parallel to the growth direction 
$z$, the first term in the right-hand side of Eq.~(\ref{current_general})
is given by\cite{footnoteMPGE}
\begin{align}
\label{noPGE}
j^{PGE}_{x}=\chi_{xxz}(B_z) E_{\omega,x} E_{\omega,z}^*+\chi_{xyz}(B_z) E_{\omega,y} E_{\omega,z}^*+\text{c.c.} \nonumber\\
j^{PGE}_{y}=\chi_{yyz}(B_z) E_{\omega,y} E_{\omega,z}^*+\chi_{yxz}(B_z) E_{\omega,x} E_{\omega,z}^*+\text{c.c.}\:,
\end{align}
where $\chi_{xxz}(B_z)$ and $\chi_{yyz}(B_z)$ are even functions of the magnetic field,
$\chi_{xyz}(B_z)$ and $\chi_{yxz}(B_z)$ are odd functions, 
and $x \parallel [1\bar{1}0]$ and $y \parallel [110]$ are the in-plane axes. 
Equations~(\ref{noPGE}) show that all  
contributions of the photogalvanic effect requires a nonzero component of the 
radiation electric field along the growth direction, $E_{\omega,z}$. Thus, the 
observation of the substantial photocurrent excited by 
$s$-polarized light ($E_{\omega,z} = 0$), 
rules out the photogalvanic effect as an origin of the observed 
resonant photocurrent (see inset in Fig.~\ref{cwtransverse}).  

Now we turn to the photon drag effect described by the second term 
in the right-hand side of  Eq.~(\ref{current_general}). 
Symmetry analysis shows that, in contrast to photogalvanics, 
photon drag effect is allowed for both $s$- and $p$-polarizations. 
In the case of $s$-polarized radiation  and the ($xz$) incidence plane,
i.e., for the  $\bm{E}_{\omega} \parallel y$, 
the current density is given by 
\begin{align}
j^{PDE}_{x}=\phi_{xxyy}(B_z) \, q_x |E_{\omega,y}|^2 \nonumber\\
j^{PDE}_{y}=\phi_{yxyy}(B_z) \, q_x |E_{\omega,y}|^2 \:,
\end{align}
where $\phi_{xxyy}(B_z)$ is an even function of the magnetic field and $\phi_{yxyy}(B_z)$ 
is an odd function. 
Consequently, we obtain that the longitudinal photocurrent ($j_x$) is even 
in the magnetic field and 
the transverse one ($j_y$) is odd. Similar results are obtained for $p$-polarized radiation. 

Polarization dependence and different parity of magnetic field dependences of 
longitudinal and 
transverse photocurrent components, together with the negligible 
 contribution of 
the photogalvanic effect,
suggest that the photocurrent is  dominated 
by the photon drag effect
whose microscopic theory is considered in the next section.

\section{Microscopic theory}

We now turn to  microscopic mechanisms responsible for the photocurrent generation.
To that end we develop a quasi-classical theory of the 
cyclotron-resonance-assisted photon drag effect. 
Such a description is relevant to our experiments where highly doped InSb/InAlSb 
quantum wells with the Fermi energy about 50\,meV are exited by terahertz radiation 
with photon energies at least an order of magnitude smaller.
In this approach, the second order response is  calculated 
in the framework of 
the Boltzmann kinetic equation for the electron distribution 
function $f(\bm{p},\bm{r}, t)$ in the momentum, coordinate, and time spaces:
\begin{equation}
\label{Boltzmann_kinetic}
\frac{\partial f}{\partial t}+\bm v \cdot \frac{\partial f}{\partial \bm r}+
e\left( \bm {E}(\bm{r},t)+\frac{1}{c}\left[ \bm v \times \bm B (\bm{r},t) \right] \right) 
\cdot  \frac{\partial f}{\partial \bm p}=\text{St}\left[ f\right],
\end{equation}
where 
$\bm{v} = \bm {p}/m^*$ is the electron velocity, 
$e$ is the electron charge, 
${\bm{B}(\bm{r},t) = \bm{B} + \bm{B}_{\omega} {\rm e}^{i\bm q \cdot \bm r -i \omega t}
+\bm{B}_{\omega}^* {\rm e}^{-i\bm q \cdot \bm r + i \omega t}}$ is the total magnetic 
field consisting of the static field $\bm B \parallel z$  and alternating magnetic field of the THz wave
with the magnitude $B_{\omega}$, 
and  $\text{St}\left[ f\right]$ is the collision integral. 

The  electric current density is given by the 
standard equation $\bm j = 2 e \sum_{\bm p} {\bm v}\, f(\bm p, \bm{r},t)$, 
where a factor of 2 accounts for  spin degeneracy. 
The  \textit{dc} photoresponse is determined by the nonequilibrium 
corrections to the 
distribution function 
which are proportional to $E_{\omega,\|} B_{\omega,z}$ and $q_{\|} E_{\omega,\|}^2$, 
where the symbol ``$\parallel$'' 
denotes the in-plane components.
The terms yield two contributions to the current:
(i) due to the dynamic Hall effect resulting from the joint action of electric and 
magnetic
fields of the radiation\cite{BARLOW1954} and (ii) due to spatial oscillations of 
the electric field
and, consequently, electron distribution\cite{perelpinskii73}.
They were consistently considered for bulk materials\cite{perelpinskii73} and 
two-dimensional systems with linear dispersion.\cite{PRL10,GlazovGanichev}
Note that the former contribution can be also rewritten via the 
photon wavevector giving $q_{\|} E_{\omega,\|}^2$, 
since the amplitudes of electric and magnetic fields in the plane wave are coupled by 
$\bm {B}_{\omega} =c [\bm {q} \times \bm{E}_\omega]/\omega$.
Application of the static magnetic field $\bm B \parallel z$, besides deflecting 
the current due to the Hall effect, leads to the drastic enhancement
of the photon drag effect at cyclotron resonance. 
Estimations for the resonant photocurrent 
in InAs/GaSb structures is given in Ref.~\onlinecite{DmitrievJETPL}.

The details of calculations are as follows.  We solve the kinetic 
equation Eq.~(\ref{Boltzmann_kinetic})
by expanding the distribution function $f(\bm{p}, \bm{r},t)$ in the Fourier 
series of frequency, angular in the
momentum space, and spatial harmonics
\begin{equation}
\label{distibution_decomposition}
f = \sum\limits_{n,m,l} f^{n,m,l}(p) \exp( - i n \omega t + i m \varphi_{\bm p} + i l \bm{q}_{\|} \cdot \bm{r})\:,
\end{equation}
where $\varphi_{\bm p}={\rm arctan} (p_y/p_x)$ is the polar angle of the vector $\bm p$. 
Note that in thermal equilibrium, the
distribution function 
is described by  the harmonic
${f^{0,0,0}(p)=\{1+\exp[(\varepsilon_p - E_F)/k_B T]\}^{-1}}$, with 
$\varepsilon_p=p^2 /(2m^*)$ being the electron energy, $E_F$ 
the Fermi energy, and $T$ the temperature, while all other harmonics vanish. 
For elastic or quasi-elastic electron scattering 
by impurities or phonons, the collision 
integral takes the form
\begin{equation}
\label{collision_integral}
\text{ St}[ f] =-\sum\limits_{n,l} \sum\limits_{m \neq 0} \frac{f^{n,m,l}(p)}{\tau_m}  \exp( - i n \omega t + i m \varphi_{\bm p} + i l \bm{q}_{\|} \cdot \bm{r})\:,  
\end{equation}
where $\tau_m$ is the relaxation time of the $m$-th angular harmonic 
of distribution function. In the Fourier series representation, 
Eq.~(\ref{Boltzmann_kinetic}) has the form of linear equation 
system for the harmonics $f^{n,m,l}(p)$
\begin{multline}
\label{equation_system}
 \Gamma^{n,m}f^{n,m,l}
- i \dfrac{m\, e}{c m^*} \left( B_{\omega} f^{n-1,m,l-1}+ B_{\omega}^* f^{n+1,m,l+1}\right)\\
+\dfrac{e \bm{E}_{\omega}}{2}\cdot\left( \bm{o}_- \hat{K}^m_- f^{n-1,m-1,l-1}+\bm{o}_+ \hat{K}^m_+ f^{n-1,m+1,l-1}\right)\\
+\dfrac{e \bm{E}^*_{\omega}}{2}\cdot\left( \bm{o}_- \hat{K}^m_- f^{n+1,m-1,l+1}+\bm{o}_+ \hat{K}^m_+ f^{n+1,m+1,l+1}\right)\\+
i \dfrac{l v \bm{q}_{\|}}{2}\left(\bm{o}_- f^{n,m-1,l}+\bm{o}_+ f^{n,m+1,l}\right)  =0\:,
\end{multline}
where $\Gamma^{n,m}=1/\tau_m-i n\omega - i m \omega_c$, $\omega_c=e B_z/(m^* c)$ 
is the cyclotron frequency, $\bm{o}_\pm=\bm{o}_x\pm i \bm{o}_y$, $\bm{o}_x$ and $\bm{o}_y$ 
are the unit vectors along $x$ and $y$, respectively, ${\hat{K}^m_\pm=\partial_p \pm (m\pm 1)/p}$. 
Straightforward calculations show that, for ${\bm{q}_{\|} \| x}$ and 
degenerate electron gas, the components of the  drag current 
density have the form
\begin{equation}
\label{total_current}
j_{x}^{\rm pd} = \operatorname{Re}(\tilde{j}) \;, \;\;\; 
j_{y}^{\rm pd} = -\operatorname{Im}(\tilde{j})\:, 
\end{equation}
where
\begin{multline}
\label{photondrag_full_current_equation}
\tilde{j}=\dfrac{ q_{\|} e^3 |E_{\omega,\|}|^2 \, n_s}{\omega m^{*2}}  
\dfrac{\tau_p}{1-i \omega_c \tau_p}  \sum_{\pm}\Biggl\{ \tau_p
\dfrac{1\pm \xi_3}{1+(\omega \pm \omega_c)^2 \tau_p^2}\\+
\dfrac{1}{2}
\dfrac{\tau_p' E_F}{1-i \omega_c \tau_p} 
\Biggl[
\left( \dfrac{\xi_1-i\xi_2}{1+ i (\pm \omega-\omega_c)\tau_p}
+\dfrac{1\pm\xi_3}{1+ i (\pm\omega+\omega_c)\tau_p}
\right)
\\   
\pm 
\dfrac{ i \omega\tau_2 (1\mp\xi_3) }{[1+i (\pm\omega - 2  \omega_c)\tau_2][1+i (\pm\omega -  \omega_c)\tau_p]} 
\Biggr]
\Biggr\}\:,
\end{multline}
$n_s=m^* E_F/(\pi \hbar^2)$ is the electron density,  $\tau_p \equiv \tau_1$ 
is momentum relaxation time,  $\tau_p' = d \tau_p(E_F)/d E_F$, 
parameters $\xi_1$, $\xi_2$, and $\xi_3$ are determined by polarization 
state of radiation and connected to the  Stokes parameters.\cite{Saleh}
The parameters are given by 
$\xi_1=(|E_{\omega,x}|^2-|E_{\omega,y}|^2)/E_{\omega,\|}^2$,  $\xi_2=(E_{\omega,x} E_{\omega,y}^*+E_{\omega,x}^* E_{\omega,y})/E_{\omega,\|}^2$,
and $\xi_3=i(E_{\omega,x} E_{\omega,y}^*-E_{\omega,x}^* E_{\omega,y})/E_{\omega,\|}^2$,  
the latter one  reflecting the radiation helicity. 
Equation~(\ref{photondrag_full_current_equation}) shows that the photon drag current
consists of two contributions: one being proportional to $\tau_p$ (first term in the 
curly brackets)
and the other to the first derivative $\tau_p'$. It is remarkable that, while the first 
contribution exhibits
the resonance at CR, i.e., for $\omega = \omega_c$, the second one has an additional 
resonance at the double frequency, $\omega = 2 \omega_c$, where the radiation absorption 
does exhibit this peculiarity.  Dependence of the drag current on the polarization 
parameters $\xi_1$ and $\xi_2$ in two-dimensional systems with parabolic energy spectrum 
arises to the extent of energy dependence of the momentum relaxation time. 
At the same time, the strong dependence of the photocurrent on radiation helicity 
results from the circular dichroism of the cyclotron-resonance absorption.

Finally, for the case of weak energy dependence of the momentum
relaxation time, i.e., $\tau'_p \ll \tau_p/E_F$, which is relevant for
short-range scattering,\cite{footnote2e} and linearly polarized
radiation, Eq.~(\ref{photondrag_full_current_equation}) reduces to
\begin{eqnarray}
j_{x}^{\rm pd} &=&   \dfrac{ q_{\|} e^3 |E_{\omega,\|}|^2 \, n_s}{\omega m^{*2}}  \dfrac{\tau_p^2}{1+(\omega_c \tau_p)^2}  
\sum\limits_{\pm}\dfrac{1}{1+(\omega \pm \omega_c)^2 \tau_p^2}\:, \nonumber  \\
j_{y}^{\rm pd} &=& -\omega_{c}\tau_{p} j_{x}^{\rm pd}\:.
\label{CRcurrents}
\end{eqnarray}
In the next section we analyze  this result, compare it with experimental data 
and show that it describes well all experimental findings.

\section{Discussion}
\label{discussion}

The above theory is developed in the assumption of a weak 
electron gas heating which is relevant to our experiments applying 
low-power radiation of the $cw$ laser. 
In these measurements, the corresponding photocurrents  have been detected 
by the voltage drops,  $U_{x,y}$, picked up across a 1\,M$\Omega$ load resistors
being much larger than the resistance of the QW structures.
To obtain $U_{x,y}$ from Eq.~(\ref{CRcurrents}) we find the steady-state distribution 
of the electric potential $\Phi(r)$
 by solving the continuity equation $\operatorname{div} \bm{j}=0$ 
with boundary conditions setting the zero electric current across 
the sample edges.
Here the total electric current, $\bm{j}$, is written in the form
\begin{equation}
\label{total_electric_current}
j_\alpha=j_\alpha^{\rm pd}-\sum\limits_{\beta=x,y} \sigma_{\alpha\beta} \nabla_\beta \Phi\:,
\end{equation}
where $j_\alpha^{\rm pd}$ is the local current density given 
by Eq.~(\ref{CRcurrents}),
the second term describes the drift current, 
$\sigma_{xx}=\sigma_{yy}=\sigma_0/[1+(\omega_c \tau_p)^2]$,   
$\sigma_{xy}=\sigma_{yx}=-\omega_c \tau_p \sigma_0/[1+(\omega_c \tau_p)^2]$ 
are the components of the conductivity tensor in a static magnetic field, 
$\sigma_{0}=n e^2 \tau_p/m^*$ is the zero-field conductivity.

\begin{figure}[h]    
\includegraphics[width=0.5\linewidth]{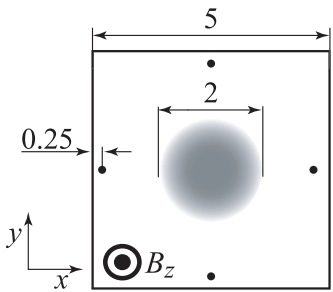}
\caption{Sample and contact dimensions in millimeters. 
Gray circle shows the laser spot located in the center of the sample.
The spot has the Gaussian profile of about $2$\,mm diameter.
Voltage drops $U_x$ and $U_y$ are measured between opposite 
contacts. 
}
\label{geometry}
\end{figure} 

\begin{figure}[tb]    
\includegraphics[width=1.0\linewidth]{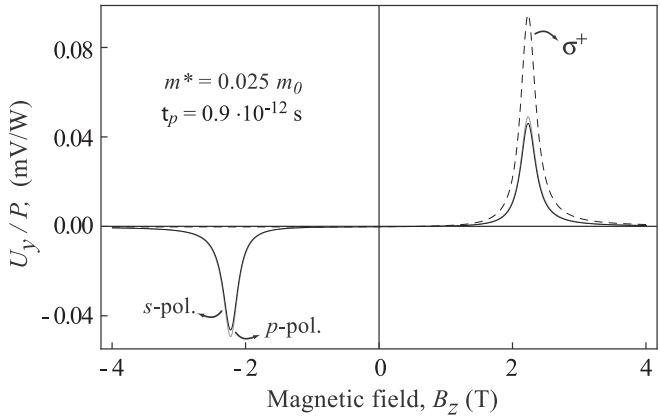}
\caption{Theoretical dependences of $U_y/P$ on magnetic field for $s$-, 
$p$- and $\sigma^+$-polarizations. 
}
\label{theory_uy}
\end{figure}

\begin{figure}[tb]    
\includegraphics[width=1.0\linewidth]{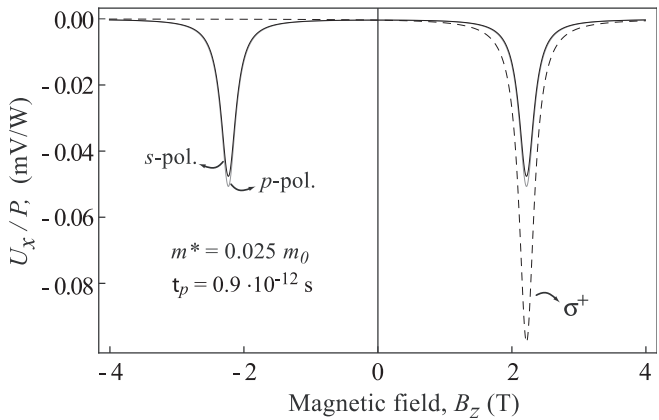}
\caption{Theoretical dependences of $U_x/P$ on magnetic field for $s$-, 
$p$- and $\sigma^+$-polarizations. 
}
\label{theory_ux}
\end{figure}

Figures~\ref{theory_uy} and~\ref{theory_ux} show calculated dependences of the
photovoltages $U_{x,y}$ on the magnetic field. The calculations are carried out
for the  momentum relaxation time, $\tau_p = 1.1\cdot 10^{-12}$~s and the 
effective mass $m^*= 0.025 \cdot m_{0}$ both obtained by the 
transmission measurements, the incidence angle $\theta_0=20^\circ$; the sample 
dimensions, the position of the contacts\cite{footnoteposition} and the Gaussian 
laser beam profile are shown in Fig.6. Comparison of the results of 
Figs.~\ref{theory_uy} and~\ref{theory_ux} 
with the corresponding experimental data shown in
Figs.~\ref{cwtransverse} and~\ref{cwparallel} demonstrates that
all essential features of the observed photocurrent are well 
described by the microscopic model of the cyclotron-resonance-assisted 
photon drag effect. In particular, for linearly polarized radiation, 
 $U_y$ and $U_x$ are described by the odd and even functions of the magnetic field. 
Estimations of the photovoltage  magnitude, 
carried out after Eqs.~(\ref{CRcurrents}) and (\ref{total_electric_current})
and using parameters determined from the transport and CR transmission measurements,
yield $U_x =0.05$\,mV/W for degenerate 2DEG, which is in good agreement with the value 
obtained in the experiments. Both the theory and experiment 
show that the photocurrent at the cyclotron resonance is enhanced by more than 50~times 
as compared to the one excited at zero field, 
Figs.~\ref{cwtransverse},~\ref{cwparallel},~\ref{theory_uy}, and~\ref{theory_ux}. 
Note, that in   another narrow band material, InAs/GaSb QWs, 
only a weak resonance with an amplitude comparable to that at zero magnetic field 
has been detected.\cite{DmitrievJETPL}

Solution of Eq. (12) with currents given by Eqs. (11) allows us to fit photovoltage data 
and obtain the momentum relaxation times which found to be
equal to  $0.9 \cdot 10^{-12}$ and $0.7 \cdot 10^{-12}$\,s for 
$T = 50$ and 80\,K, respectively. The former value and the reduction of the
relaxation time with the temperature increase are in a good agreement with the 
magneto-transport measurements. Moreover, our data provide even deeper insight in
the scattering mechanisms. Indeed, the experimental data are well described by 
the calculations  after Eqs.~(\ref{CRcurrents}) and (\ref{total_electric_current}) 
obtained assuming a weak energy dependence of the momentum relaxation time, i.e., $\tau'_p \ll \tau_p/E_F$.
The validity of this assumption manifests itself in 
the magnetic field dependence of the photocurrent, see Figs.~\ref{cwtransverse} and~\ref{cwparallel}, 
which does not show any peculiarities at  $B = B_c / 2$ at which for the opposite unequality
an additional resonance should appear, see Eqs.~(\ref{photondrag_full_current_equation}).

Finally, the calculations show that the amplitudes of the resonance are nearly the  same
for $s$- and $p$-polarized radiation, which agrees well 
with experimental findings, see Figs.~\ref{cwtransverse} and~\ref{cwparallel}. 
Indeed, the polarization dependence of the photon drag current for energy-independent 
relaxation time is described by 
$|E_{\omega,\|}|^2$, see Eqs.~(\ref{CRcurrents}), which  
for $s$- and $p$-polarizations are given by $|E_{\omega,\|}|^2=2\pi t_s^2 I_0 /c$ and 
$|E_{\omega,\|}|^2=2\pi t_p^2 I_0 \cos^2 \theta/c$, respectively. 
Here, $t_s$ and $t_p$ are the amplitude transmission coefficients for $s$- and $p$-waves,
$I_0$ is the radiation intensity, $\theta$ is the angle of refraction 
related with the incidence angle $\theta_0$ by 
$ \sin \theta=\sin \theta_0 / n_{\omega}$, and $n_{\omega}$ is the refraction index. 
Consequently, for small angles of incidence with $\cos \theta \approx 1$ 
and $t_p \approx t_s$, the fields $|E_{\omega,\|}|$ and the resulting 
photocurrents are close to each other for the considered polarization states. 
Figures~\ref{theory_uy} and~\ref{theory_ux} also show the  
photocurrent excited by right-handed, $\sigma^+$, circularly polarized radiation. 
It is seen that, as expected for CR, here the resonant signal is generated for one 
magnetic field polarity only (positive $B_z$) and that its magnitude is enhanced by 
factor~2 compared to that excited by linearly polarized radiation.

Now we briefly address the observed oscillations of the photocurrent. The
oscillations have been detected at low temperature and applying $cw$ laser 
radiation only. They correlate with the Shubnikov -- de Haas oscillations and 
vanish at high  lattice or electron temperature. 
The latter has been proved applying pulsed high power radiation resulting in a strong electron gas heating,
which manifests itself in the substantial photoconductive signal.\cite{GanichevPrettl}
In this particular case,  even at $T=4.2$\,K we only detected  the resonant photocurrent. 
The correlation with SdH indicates that the resonances are caused by 
periodic modification of the electron scattering, however
the detailed mechanism of the oscillating photocurrent remains unclear and 
requires further  experimental and theoretical studies. In particular, the observed 
two times difference in the oscillation frequencies should be explored.

\section{Summary}
\label{summary}

Our observations demonstrate that in InSb/InAlSb QWs
the photon drag effect resulting from the  transfer of 
photon momentum to free electrons is 
strong  under cyclotron resonance conditions.
The microscopic theory of the photon drag effect in two-dimensional 
electron systems with parabolic energy spectrum developed 
in the framework of Newton equations
of motion and Boltzmann kinetic equation for the distribution function
describes well all experimental findings.
The observed effect provides complimentary method to study 
details of the band structure and carrier dynamic in
InSb-based two dimensional structures. In particular, 
almost identical Lorentz profiles and the resonance positions 
observed in the experiments on photocurrent excited by low power 
THz excitation and radiation transmission, demonstrate, that 
the photon drag effect can be applied for measuring 
of the effective mass. Since a large photoresponse is observed at
high temperatures and should also yield well detected signals 
for the structures with a low carrier density, this method can 
be used even under the conditions where transmission signals are unresolvable.
Furthermore, our analysis shows that investigating the 
resonance photocurrent at magnetic fields equal to a half of $B_c$ 
one can study the energy dependence of the momentum
relaxation time, gaining information not easily accessible by other experimental methods. 

\acknowledgments

The financial support from the DFG (SFB~689), the
Linkage Grant of IB of BMBF at DLR,  and RFBR, is gratefully
acknowledged.


\begin{thebibliography}{99}
\bibitem{footnote1} Note that at normal incidence the signal is not detected for any magnetic field strength.
\bibitem{footnoteMPGE} Note that generally speaking in the presence of magnetic field 
the photocurrent may also be generated by magneto gyrotropic photogalvanic 
effect.\cite{Belkov05} However, for the point group C$_{2v}$ this 
effect vanishes for $\bm{B} \parallel z$~[\onlinecite{BelkovGanichev}] 
and therefore is disregarded in the further consideration.
\bibitem{footnote2e} The absence of a noticeable energy dependence of
the relaxation time $\tau_p$ is additionally supported by the fact
that no resonant photocurrent is observed at $\omega=2\omega_c$ which
is expected from Eq.~(\ref{photondrag_full_current_equation})
elsewise.
\bibitem{footnoteposition} Note that while the longitudinal voltage, $U_x$, 
is nearly independent of distance between contact and sample 
edge the transverse one, $U_y$, is sensitive to the contact positions.
\end{thebibliography}
\end{document}